\documentclass[]{spie}  

\usepackage{amsmath,amsfonts,amssymb}
\usepackage{graphicx}
\usepackage[colorlinks=true, allcolors=blue]{hyperref}
\usepackage{float}
\title{Implementation and Characterization of the Vector Vortex Coronagraph on the SEAL Testbed}

\author[1]{Ashai Moreno}
\author[1]{Vincent Chambouleyron}
\author[1]{Rebecca M. Jensen-Clem}
\author[1]{Daren Dillon}
\author[1]{Philip M. Hinz}
\author[1]{Bruce Macintosh}
\affil[1]{Department of Astronomy and Astrophysics, University of California, Santa Cruz, CA 95064, USA}

\authorinfo{Further author information: Send correspondence to Ashai Moreno. E-mail: asanmore@ucsc.edu}

\begin{document} 
\maketitle

\begin{abstract}
The Santa Cruz Extreme AO Lab (SEAL) testbed is an optical bench meant to design and develop new wavefront control techniques for high-contrast imaging for segmented telescopes. These techniques allow for astronomical efficiency in exoplanet imaging and characterization. SEAL consists of several wavefront sensors (WFS) and deformable mirrors (DM) that are currently performing techniques like predictive control or non-linear reconstruction. In this paper, we present the implementation and characterization of a new coronagraphic branch on SEAL and assess the contrast limitations in the testbed. For our coronagraphic branch, we used a vector vortex coronagraph which has high contrast performance. The W. M. Keck Observatory also uses a vortex coronagraph, allowing us to compare the limitations with our own coronagraph. We relied on the testbed and simulations of the vortex coronagraph to compare performance with expected ones. To create a more reliable simulation, we also injected in our numerical model data collected by a Zernike Wavefront sensor (ZWFS) used to perform fine wavefront sensing on the bench. Now that the coronagraphic branch is aligned on SEAL, we will be able to use contrast as a metric for the performance of wavefront control methods on the bench.
\end{abstract}


\section{INTRODUCTION}
\label{sec:intro}  

The detection and characterization of exoplanets is one of the main tasks of modern astronomy. It allows for a range of understanding, from planet formation to the possibility of life on other worlds. The method of direct imaging gives us a comprehension of these systems but is hard to execute due to atmospheric turbulence as well as the fact that exoplanets are usually hidden due to the glare of the star they orbit. Adaptive optics (AO) and optical instruments known as coronagraphs are meant to combat these problems by flattening distorted wavefronts and creating high contrasts between bright and dim objects close to each other by deflecting the light from the bright objects. Vortex coronagraphs are a type of coronagraph that have proven to perform high-performance starlight ejection~\cite{Mawet:09}, with the ability to work under broadband light. Unlike a Lyot coronagraph, a vortex coronagraph is able to diffract starlight away from the center of the pupil using a focal plane mask consisting of a phase ramp~\cite{doelman2023laboratory}.

In this paper, we present the implementation of a coronagraphic branch on the SEAL testbed using a vector vortex coronagraph (VVC) and the performance assessment using simulations as well as real data from SEAL. In section 2, the SEAL testbed and the implementation of the vector vortex coronagraphic branch are presented. In section 3, we discuss the performance characterization by comparing SEAL coronagraphic performance with what was expected through simulation. In section 4, the SEAL testbed stability is studied in both open and closed loop using contrast as the metric. Finally, in section 5 we provide conclusions and perspectives. 

\section{Implementation of Vector Vortex on SEAL Testbed}

\subsection{SEAL Testbed}

SEAL~\cite{jensenclem2021santa} is designed to develop new wavefront sensing techniques and control strategies dedicated to high-contrast imaging for segmented ground-based telescopes. In particular, SEAL was designed to have strong similarities with the high-contrast branch of the W.M. Keck Observatory, allowing for research and development in technologies and concepts that could later be implemented on sky. 

\begin{figure} [H]
   \begin{center}
   \begin{tabular}{c} 
   \includegraphics[scale=0.4]{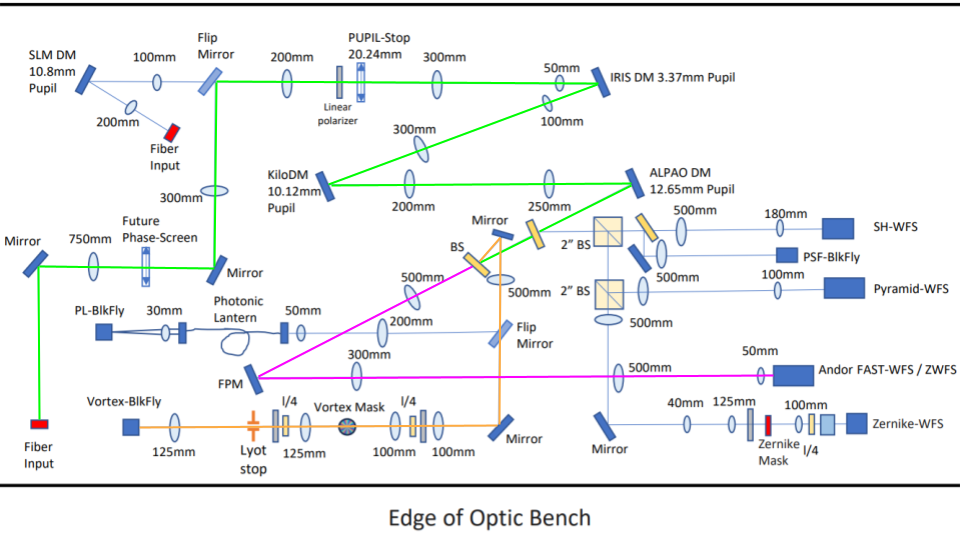}
   \end{tabular}
   \end{center}
   \caption[example] 
   { \label{fig:layout_path} 
Schematic of the SEAL testbed layout. The paths that we are using in this paper are highlighted in green (common path for ZWFS and VVC), orange (path for VVC), and purple (path for ZWFS).}
   \end{figure}

SEAL consists of a segmented deformable mirror (DM) (IRISAO) composed of 37 segments in the aperture, a low order DM with 8 actuators across pupil (ALPAO), and a high order DM with 24 actuators across pupil (BMC). Our optical bench (Figure~\ref{fig:layout_path}) is working at 635 nm and also possesses a spatial light modulator (SLM), to emulate atmospheric turbulence, and multiple wavefront sensors (WFS) such as a Zernike WFS (ZWFS) and a Pyramid WFS (PWFS). In order to mimic the Keck high contrast branch, we decided to implement a coronagraphic branch onto SEAL with a vortex coronagraph, similar to the setup used on Keck-NIRC2\cite{xuan2018characterizing}.   

\subsection{Vector Vortex Coronagraph}

\begin{figure} [H]
   \begin{center}
   \begin{tabular}{c} 
   \includegraphics[scale=0.15]{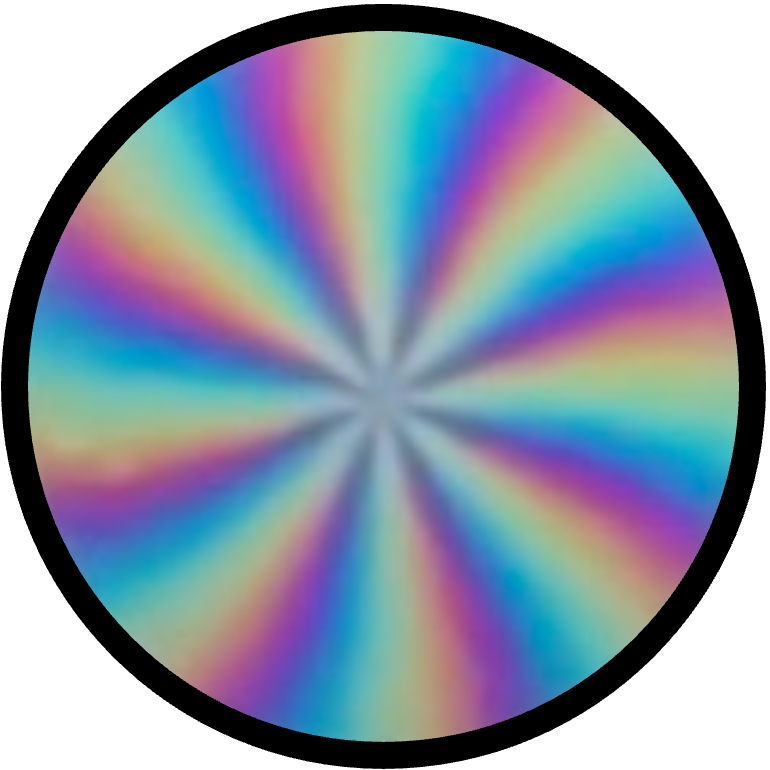}
   \end{tabular}
   \end{center}
   \caption[example] 
   { \label{fig:vortex_mask} 
Charge four vector vortex mask over polarized light.}
   \end{figure}

The coronagraphic branch on SEAL consists of a vector vortex mask (Figure \ref{fig:vortex_mask}), designed to keep good performance in broadband light by implementing "geometrical" phase shifts independent of wavelength. These phase shifts are created by liquid crystals that works on circular polarization and that are known to exhibit a leakage term~\cite{10.1117/12.858240}, which needs to be mitigated by using a system of polarizers leading to a loss of 50\% of the light.
The branch also includes a diaphragm acting as our Lyot stop (circular), a blackfly camera for focal plane imaging, and two sets of circular polarizers/analysers (each being composed of a quarter-wave plate and a linear polarizer). The optical design is a follows: light goes through a first lens which creates a collimated beam and then reaches the first circular polarizer to create a right-handed circular polarized light. The beam then goes through another lens, converging onto the vector vortex mask. The vector vortex mask used in this experiment is of charge 4, where the relation of charge (\(l\)) and the phase ramp of our mask complex amplitude (\(m\)) is the following:
\[m(x,y) = l\theta(x,y)\]
\(\theta\) in this case represents the azimuthal coordinate on our mask. The charge represents the amount of times our mask goes through a full phase shift of 2\(\pi\). After the mask, the light goes through another lens, bringing it back into collimated space where it faces another set of QWP and linear polarizer to eliminate the left-hand leakage created by the vortex mask. The light rejected outside the pupil is then blocked by the Lyot stop. Our Lyot stop has a radius of 79\% compared to the radius of the SEAL pupil. We chose this size because it was able to be easily reproduced when opening and closing the diaphragm during realignments. Finally, the light converges back into the focal plane and onto the camera using the last lens. Our branch also includes a pupil plane viewing mode on the same camera. This mode is enabled by sliding a lens in the beam after the Lyot stop plane. This capability allows us to align our vector vortex mask and to measure the size of our Lyot stop relative to the pupil.

\begin{figure} [H]
   \begin{center}
   \begin{tabular}{c} 
   \includegraphics[scale=0.4]{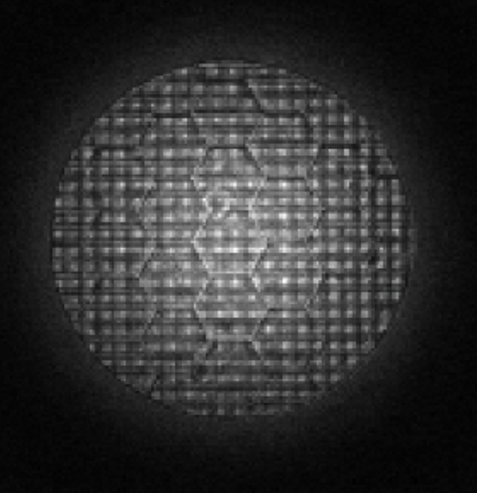}
   \end{tabular}
   \end{center}
   \caption[example] 
   { \label{fig:zwfs} 
Data collected from the reflective scalar Zernike WFS for a flat wavefront.}
   \end{figure}

The setup uses a reflective ZWFS~\cite{2023SPIE12680E..02C} (Figure~\ref{fig:zwfs}) which is a highly sensitive wavefront sensor located on a separate branch, but which is coaligned with the coronagraphic branch, being therefore used as the reference for our experiment. This sensor was used as the reference one since it is the one with the least non-common path aberrations (NCPAs) for our coronagraphic data (see Figure \ref{fig:layout_path}). Once the mask is aligned, this setup allows performance of a closed-loop on the reflective ZWFS to center the SEAL PSF on the vortex.

\section{Performance characterization}
\label{sec:Performance characterization}

The goal for this section is to characterize our performance for the coronagraphic branch on SEAL. In order to do so, we created a simulation of our vector vortex coronagraph and compared it to our performance on the SEAL testbed. 

\subsection{Expected Performance from Simulations}

\begin{figure} [H]
   \begin{center}
   \begin{tabular}{c} 
   \includegraphics[scale=.6]{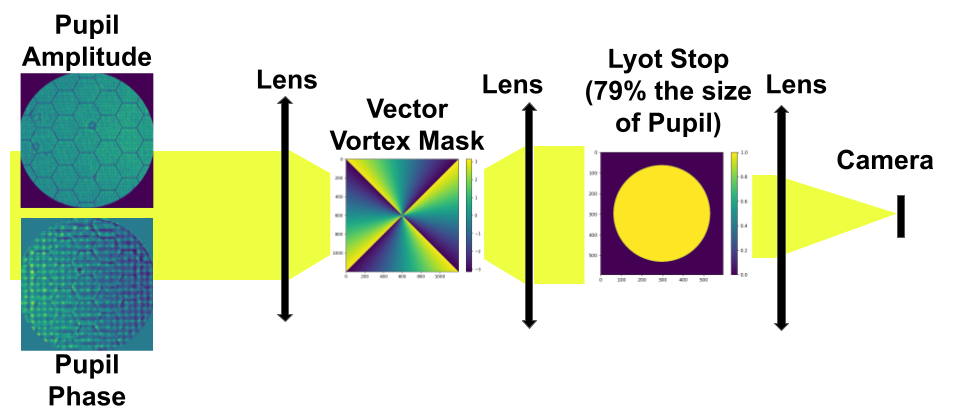}
   \end{tabular}
   \end{center}
   \caption[example] 
   { \label{fig:simulated_components} 
Figure of our simulated components for our coronagraphic setup.}
   \end{figure}

We built a simple model of the VVC and injected data collected from SEAL. The data injected corresponds to the phase and amplitude in the pupil plane (see Figure \ref{fig:simulated_components}), measured by the ZWFS (amplitude can simply be measured by taking an image off mask). Segments gap from IRISAO DM can be seen in the amplitude data, while well-known effect of "quilting" on BMC DM is clearly visible on estimated phase by the ZWFS. Both of these effects have an impact on the coronagraphic PSF.

Our lenses were simulated using Fraunhofer propagation. Fraunhofer propagation uses Fourier transforms to simulate propagation. This allowed us to also simulate the diffraction pattern caused by our segmented aperture. The vector vortex mask was simulated with a charge of 4 (corresponding to the charge of the one used experimentally). Our simulated Lyot stop is simply defined as a circular aperture with a radius 79\% the size of our simulated pupil (Figure \ref{fig:simulated_components}).

\begin{figure}[H]
   \begin{center}
   \begin{tabular}{c} 
   \includegraphics[height=5cm]{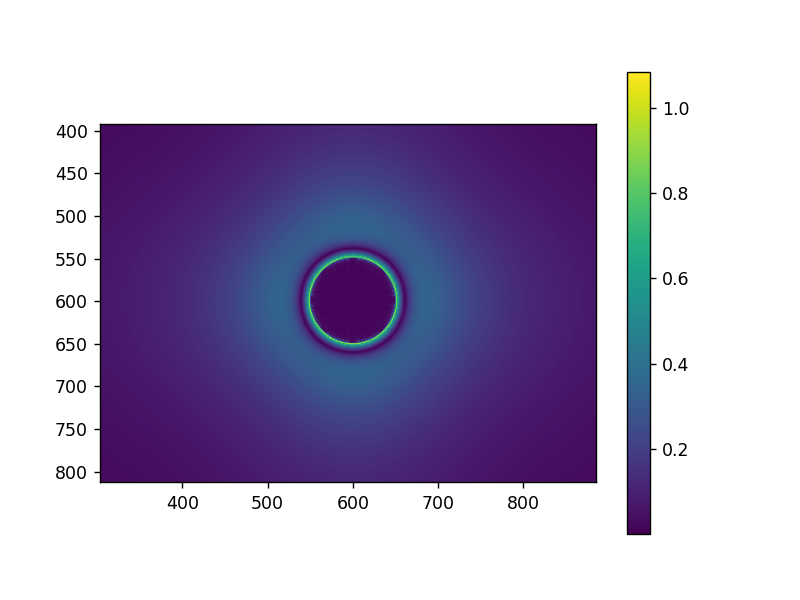}
   \includegraphics[height=5cm]{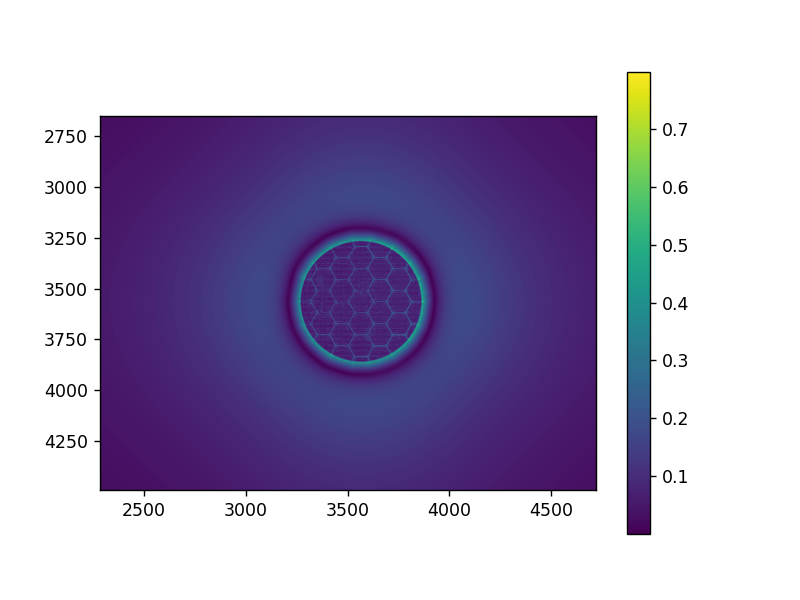}
   \end{tabular}
   \end{center}
   \caption[example] 
   { \label{fig:sim_fpm_effects} 
Simulated effects of the vector vortex mask in the Lyot stop plane with a circular (left) and segmented aperture (right).}
   \end{figure}

\begin{figure} [H]
   \begin{center}
   \begin{tabular}{c} 
   \includegraphics[height=5cm]{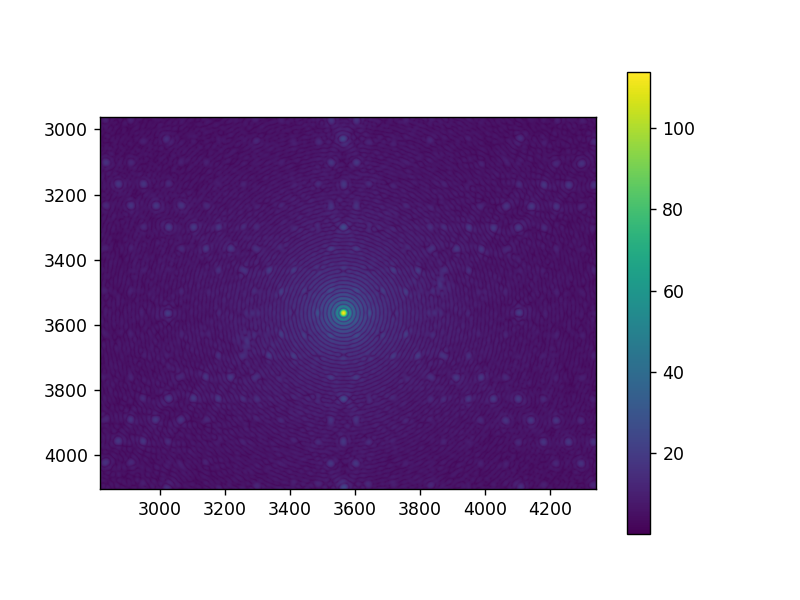}
   \includegraphics[height=5cm]{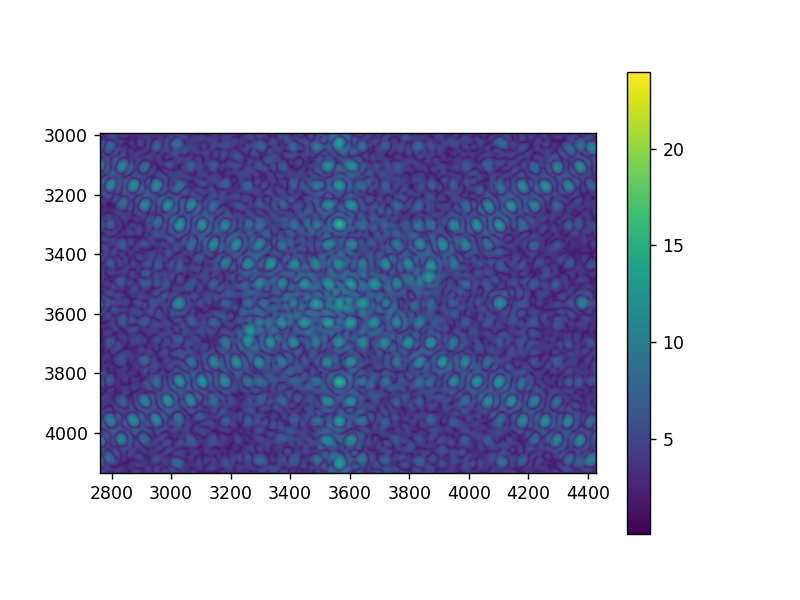}
   \end{tabular}
   \end{center}
   \caption[example] 
   { \label{fig:sim_psfs} 
Simulated non-coronagraphic (left) and coronagraphic (right) PSFs.}
   \end{figure}

The effects of our vector vortex mask in the Lyot stop plane were observed using the simulation with both a full circular aperture and a segmented aperture (Figure \ref{fig:sim_fpm_effects}). Comparing the effects of both apertures, we can see that gaps between the segments lead to starlight leakage, compared to a circular aperture where all the light seems to be rejected from the center. Due to the diffraction of light leaking through the gaps, our PSFs have highly structured speckles which are especially prominent in the coronagraphic simulation, where that is one of the main sources of starlight leakage (Figure \ref{fig:sim_psfs}).

This simulated setup also resulted in a coronagraphic point spread function (PSF) and a non-coronagraphic PSF by removing the simulated Lyot stop and vector vortex mask (Figure \ref{fig:sim_psfs}). Impacts of the segment gaps are clearly visible in the coronagraphic image.

\begin{figure} [H]
   \begin{center}
   \begin{tabular}{c} 
   \includegraphics[scale=.3]{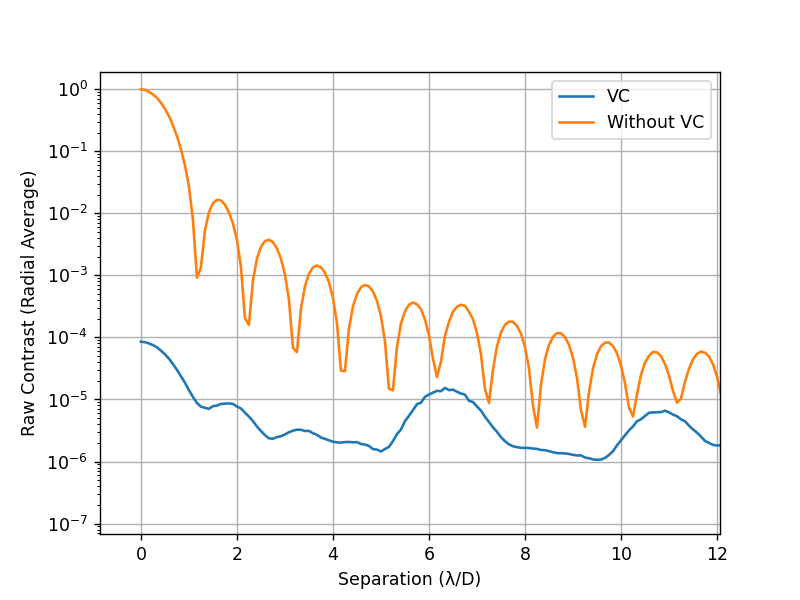}
   \end{tabular}
   \end{center}
   \caption[example] 
   { \label{fig:simulated_raw} 
Simulated raw contrast curve with a Lyot stop size of 79\% the size of the pupil.}
   \end{figure}

Using equation (\ref{eq:raw_contrast}), where \(R(x,y)\) is our coronagraphic PSF and \(max[PSF_{star}(x,y)]\) is the maximum value of our non-coronagraphic PSF\cite{jensen2017new}, we were able to compute what we define in this paper as the raw contrast of our simulated system (Figure~\ref{fig:simulated_raw}). From simulated images given in Figure \ref{fig:sim_psfs}, we measured the the raw-contrast radial average between separations of 0 to 12 \(\lambda/D\). An average raw contrast of about \(5 \times 10^{-6}\) is observed between 2 to 4 \(\lambda/D\).
\begin{equation}
\label{eq:raw_contrast}
raw\, contrast = \frac{R(x,y)}{max[PSF_{star}(x,y)]}
\end{equation}

\subsection{SEAL Performance}

\begin{figure} [H]
   \begin{center}
   \begin{tabular}{c} 
   \includegraphics[height=5cm]{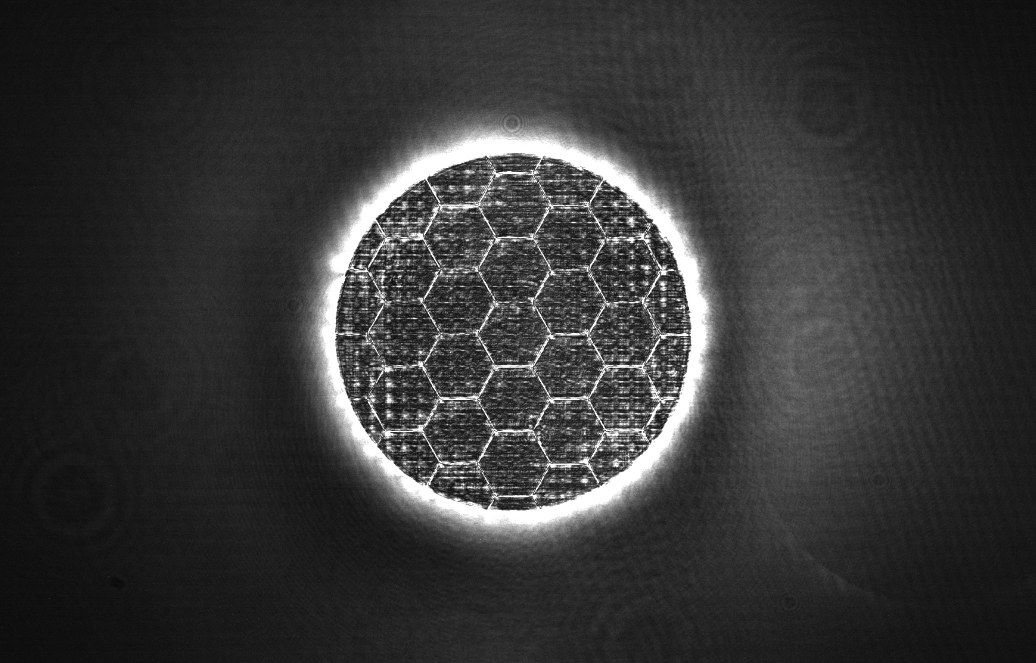}
   \end{tabular}
   \end{center}
   \caption[example] 
   { \label{fig:real_fpm_effects} 
Effects of the vector vortex mask in the Lyot stop plane on SEAL.}
   \end{figure}

\begin{figure} [H]
   \begin{center}
   \begin{tabular}{c} 
   \includegraphics[height=5cm]{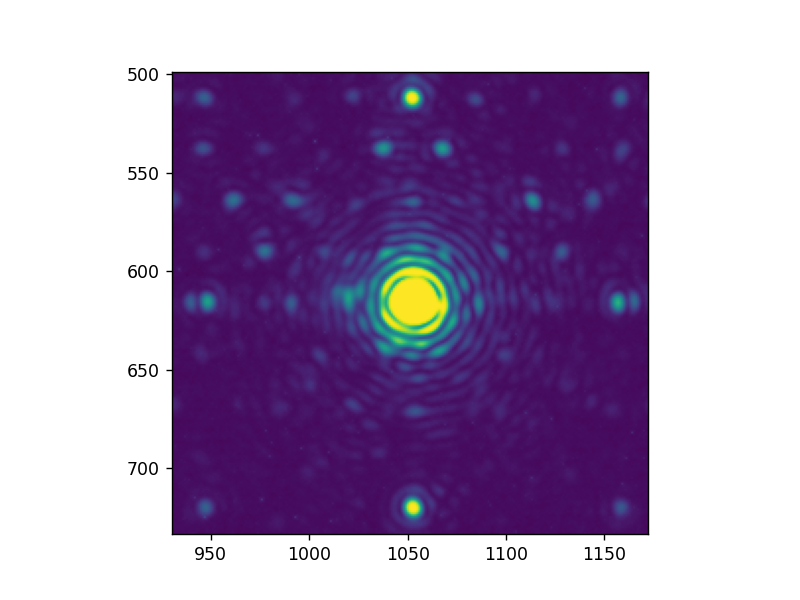}
   \includegraphics[height=5cm]{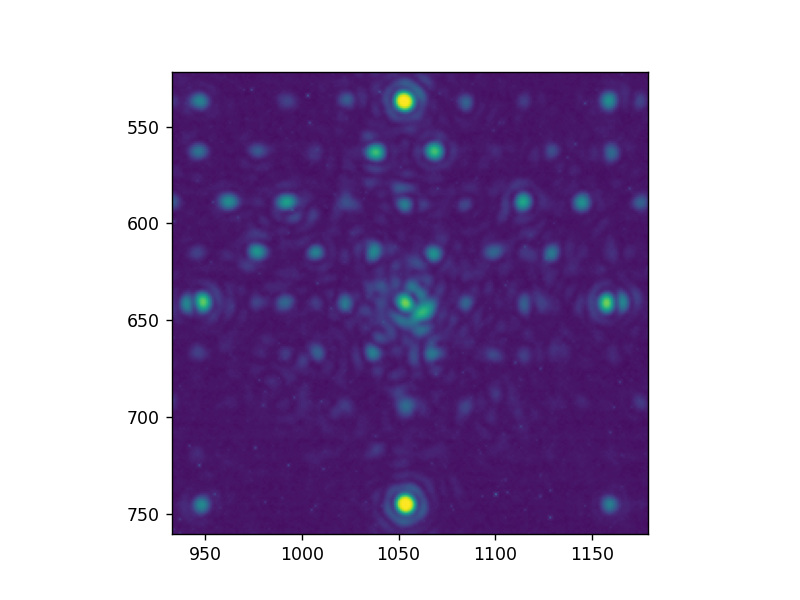}
   \end{tabular}
   \end{center}
   \caption[example] 
   { \label{fig:real_psfs} 
SEAL non-coronagraphic (left) and coronagraphic (right) PSFs.}
   \end{figure}

Coronagraphic and non-coronagraphic PSFs (Figure~\ref{fig:real_psfs}) were obtained on the SEAL testbed. To do so, the loop was closed on the ZWFS to correct for any static aberrations. From there, our vector vortex mask was aligned by maximizing rejected light in the Lyot stop plane, using the pupil viewing mode (with the Lyot stop opened). The effects of the vector vortex mask in the Lyot stop plane were recorded (Figure~\ref{fig:real_fpm_effects}) to be compared with the simulation presented in the previous section \ref{fig:sim_fpm_effects}. The obtained pupil image matches the one from simulation, with a clear light leakage from segment gaps. 

The pupil viewing mode was also used to reduce the Lyot stop aperture to the desired size after centering the mask. In order to measure the non-coronagraphic PSF, both the mask and the Lyot stop were removed. The coronagraphic PSF exhibits the expected speckles coming from segment gaps (Figure \ref{fig:real_fpm_effects}). However, one can also notice a central residual PSF core in the coronographic PSF. This residual PSF most likely comes from the polarization leakage term not being entirely removed by the set of circular polarizer and analyser.

\begin{figure} [H]
   \begin{center}
   \begin{tabular}{c} 
   \includegraphics[scale=.4]{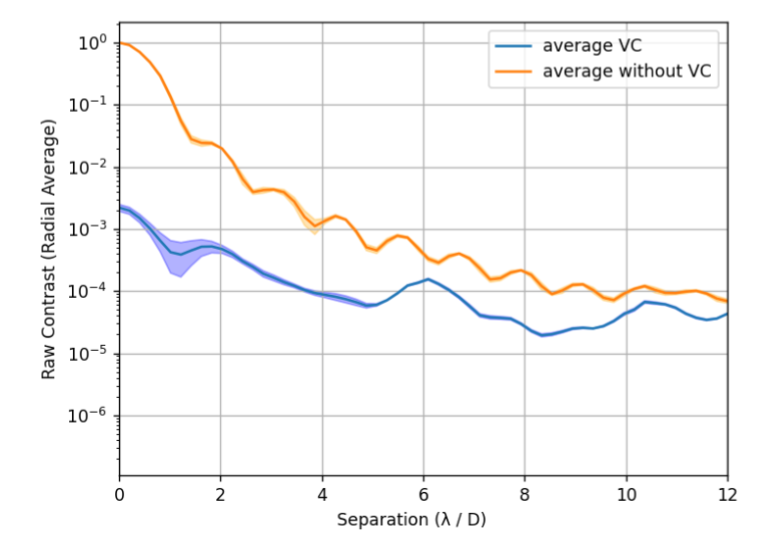}
   \end{tabular}
   \end{center}
   \caption[example] 
   { \label{fig:raw_testbed} 
Figure of our average raw contrast curve on the SEAL testbed. The lines represent the mean of the sample that was taken and the shaded regions represent the standard deviation of the sample.}
   \end{figure}

To calculate raw contrast on the testbed, the procedure mentioned at the beginning of this subsection for mask and Lyot stop alignment and closing the loop on the ZWFS was repeated several times (5 times) to check repeatability of the alignment procedure. The mean raw contrast of these separate alignments and the standard deviation of this sample was then calculated. Using this testbed data and equation (\ref{eq:raw_contrast}) we calculated our testbed raw contrast (Figure~\ref{fig:raw_testbed}). Our standard deviation and mean suggest that our contrast is robust with respect to our alignment procedure. Compared to our simulated raw contrast (Figure~\ref{fig:simulated_raw}), SEAL performance is under-performing, suggesting that we are limited by our vector vortex mask leakage term and polarizer efficiency. An average raw contrast of about \(2 \times 10^{-4}\) is observed between 2 to 4 \(\lambda/D\), about 2 orders of magnitude higher than that of our simulation, resulting in a lower contrast.

\begin{figure} [H]
   \begin{center}
   \begin{tabular}{c} 
   \includegraphics[scale=.4]{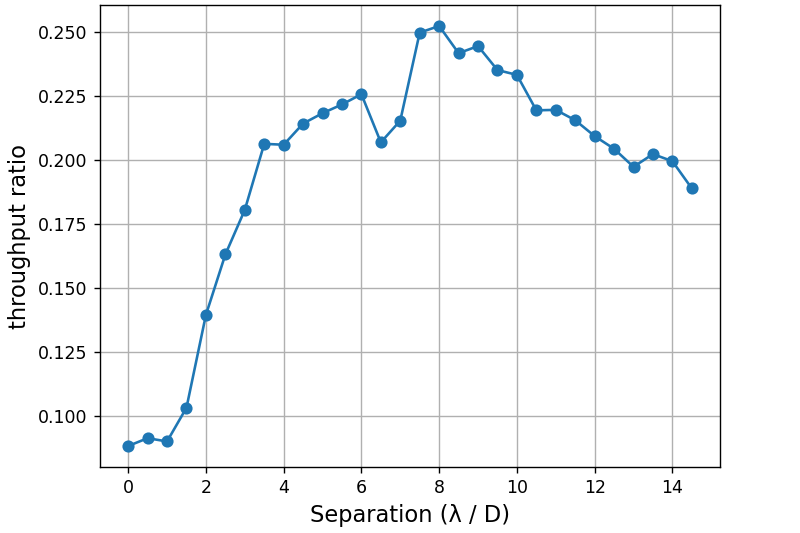}
   \end{tabular}
   \end{center}
   \caption[example] 
   { \label{fig:throughput} 
SEAL VVC throughput curve experimentally measured on the bench.}
   \end{figure}

To push the understanding of the VVC behavior, throughput was measured for separations ranging from 0 to 12 \(\lambda /D\) by applying a tip command to the high order DM to move the PSF at different separations. The tip aberrations created by the DM are not perfect for large amplitudes, leading to extra PSF aberrations. These aberrations lead to a decrease in estimated throughput for larger aberrations, which was partially compensated by calibrating the effect on the non-coronagraphic PSF. To calculate the non-coronagraphic PSF flux, the linear polarizers, QWPs, vector vortex mask, and Lyot stop were removed from the path. The throughput ratio at the given separations was computed to give us Figure~\ref{fig:throughput}. We observed a maximum throughput ratio of about 0.25. This value is close to the theoretical value (\(\approx 0.30\)) if you were to consider the amount of light being rejected by the Lyot stop size (\(\approx 40\%\)) and half of the light that is loss due to circular polarization rejection. 

\section{SEAL Contrast Evolution}
\label{sec:SEAL Contrast Evolution}

\subsection{Raw Contrast to Contrast}
Thanks to the throughput measurement of our system for different separations, it was possible to inject it for a more extensive definition of contrast~\cite{jensen2017new} which is closer from a detection limit than the raw contrast metric used until now (equation~\ref{eq:contrast}). 
\begin{equation}
\label{eq:contrast}
contrast = \frac{factor\times noise}{stellar\, aperture\, photometry}\times \frac{1}{throughput}
\end{equation}

To calculate our noise term, a ring of circular elements with diameters of 1 \(\lambda/D\) was created for each separation of our coronagraphic PSF. The mean value in each element was then calculated and used to find the standard deviation (\(\sigma\)) of noise at a given separation. Our stellar aperture photometry term was calculated with the mean value inside a similar circular element at the center of the non-coronagraphic PSF.

\begin{figure} [H]
   \begin{center}
   \begin{tabular}{c} 
   \includegraphics[scale=.4]{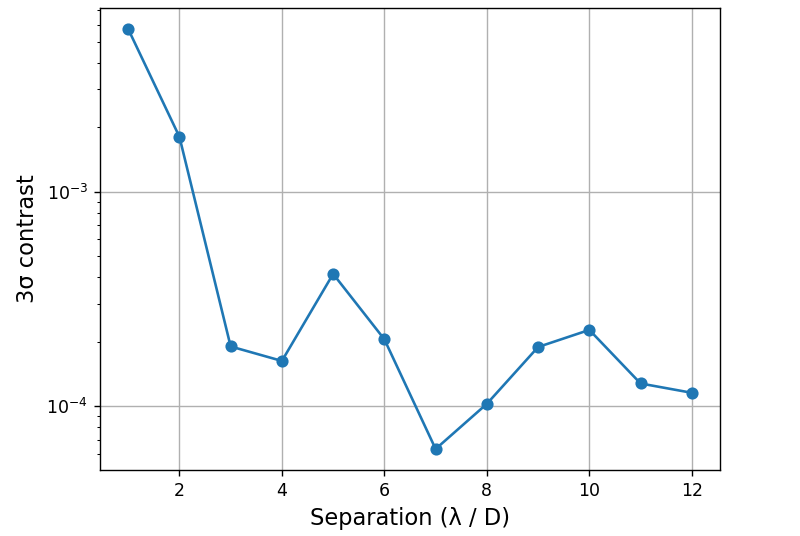}
   \end{tabular}
   \end{center}
   \caption[example] 
   { \label{fig:contrast_curve} 
Figure of our 3\(\sigma\) contrast curve. }
   \end{figure}

For our factor term in equation \ref{eq:contrast}, we chose a factor of 3. This factor was chosen because if we were to assume a Gaussian noise, anything past 3\(\sigma\) only represents 0.001 of the noise distribution. In the context of a coronagraph and its scientific purpose of detecting signals from exoplanets, our factor of 3 means that any signal that is 3 or more \(\sigma\) away from the center of the noise distribution is considered a possible detection. Our false positive fraction is minimized to 0.001 while our true positive fraction is at 0.5 if the mean of the signal detected is located at 3\(\sigma\). This factor is what could be considered as a good trade-off between the two. Using all these methods and parameters, we are able to understand what would be our detectability levels for signals 3\(\sigma\) or more from the center of the noise distribution in terms of contrast (Figure~\ref{fig:contrast_curve}).

\subsection{Time Evolution}
\begin{figure} [H]
   \begin{center}
   \begin{tabular}{c} 
   \includegraphics[scale=.3]{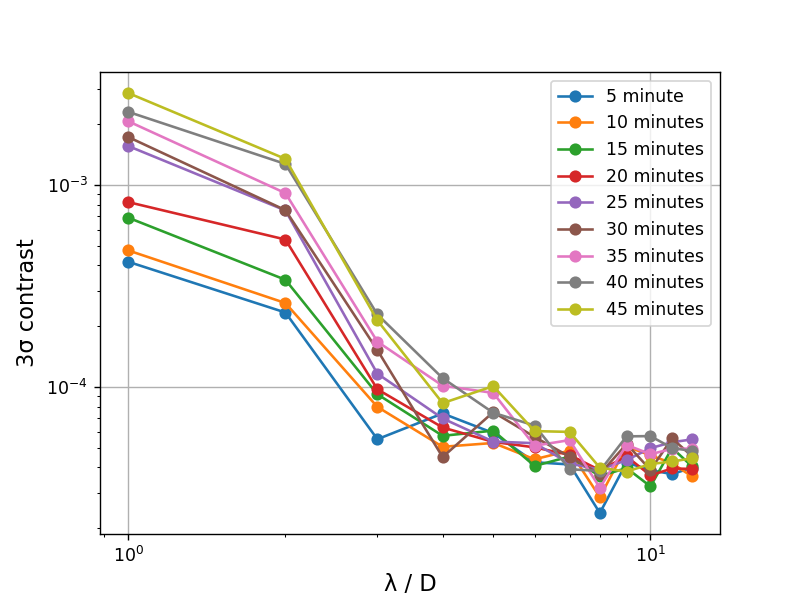}
   \includegraphics[scale=.3]{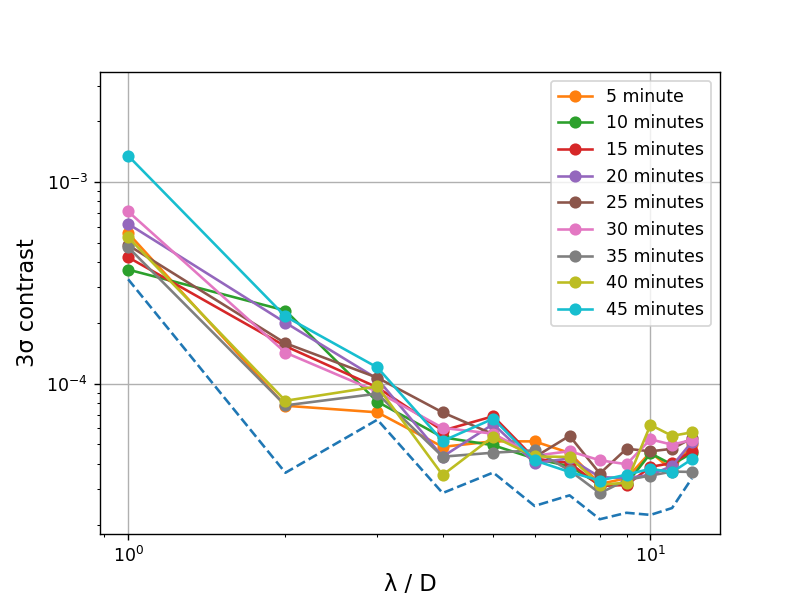}
   \end{tabular}
   \end{center}
   \caption[example] 
   { \label{fig:openloop} 
Testbed stability with the ZWFS in open-loop (left) and closed-loop (right) through time with each curve being a 5-minute step for a 45-minute interval. For our closed loop figure, the dashed line represents our noise floor which was recorded at a very small time \(\epsilon\).}
   \end{figure}

The goal is then to study SEAL stability using contrast, as defined in the previous section, for our metric. In order to do so, a coronagraphic PSF at a start time \(t_{0}\) was recorded, followed by PSFs at later times \(t_{i}\) in 5-minutes intervals during a total time of 45 minutes. Each PSF at \(t_{i}\) was subtracted by the PSF at \(t_{0}\) and the contrast for the output differential PSF was computed (equation~\ref{eq:contrast_evo}), similar to what a reference or angular differential imaging processing~\cite{Lafreniere_2007} would produce. Contrast evolution through time was recorded for two cases: (i) the loop is closed on the ZWFS until reaching a best flat, and then the loop is set to open for the full 45 minutes of measurements (Figure~\ref{fig:openloop}-left) (ii) Closed loop is maintained throughout all 45 minutes at a rate of about ~1Hz. (Figure~\ref{fig:openloop}-right). The noise floor for our closed loop test was measured by recording a coronagraphic PSF directly after recording the initial PSF at \(t = t_{0}\): in such a short amount of time, it is possible to consider that we are in a regime limited only by camera read-out-noise and photon noise when measuring the contrast. Hence, any extra contrast degradation between 2 frames can be attributed to phase instabilities.

\begin{equation}
\label{eq:contrast_evo}
contrast(t_{i}) = contrast(PSF_{t_{i}}-PSF_{t_{0}})
\end{equation}

Observing Figure~\ref{fig:openloop}, we can see that in open loop our contrast gets worse through time, while close loop resulted in a relatively stable contrast, much closer to our noise floor contrast evolution. Evolution of the closed-loop contrast is driven by evolving non-common path aberrations between the ZWFS path and the VVC branch.

\section{Conclusions and Perspectives}
\label{sec:Conclusion and Perspectives}
SEAL, a high-contrast testbed meant to test wavefront sensing and control, now has coronagraphic capabilities with the implementation of the VVC. A simulation was created to compare expected performances with true ones. In order to make the simulation more reliable, we injected data from the ZWFS and were able to see the effects in both the Lyot stop plane and focal plane. The effects of the vector vortex mask were similar for both simulation and experimental data, as light was leaking through the center due to the gaps between our segmented aperture. Our raw contrast in simulation performed better than our raw contrast on our optical bench. SEAL's under-performance can be explained by the leakage term not being entirely removed and wavefront drift through time that can be seen even in case of close loop (NCPAs).  

The throughput of our system was then measured in order to better understand the VVC installed on SEAL. With this throughput, we were able to transition our metric of performance from raw contrast to contrast. This new definition also allowed us to take into account our noise-to-signal ratio. We now have the tools needed to measure the contrast of our system. 

The time evolution of our optical bench was also studied. We observed that in open loop, our contrast slowly worsened while closed loop allowed for better stability, but still limited by NCPA drifts. 

\begin{figure} [H]
   \begin{center}
   \begin{tabular}{c} 
   \includegraphics[scale=.3]{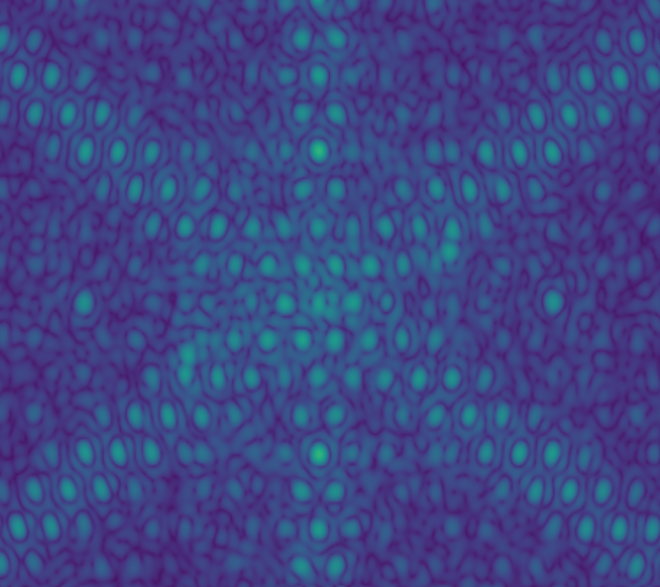}
   \includegraphics[scale=.1222]{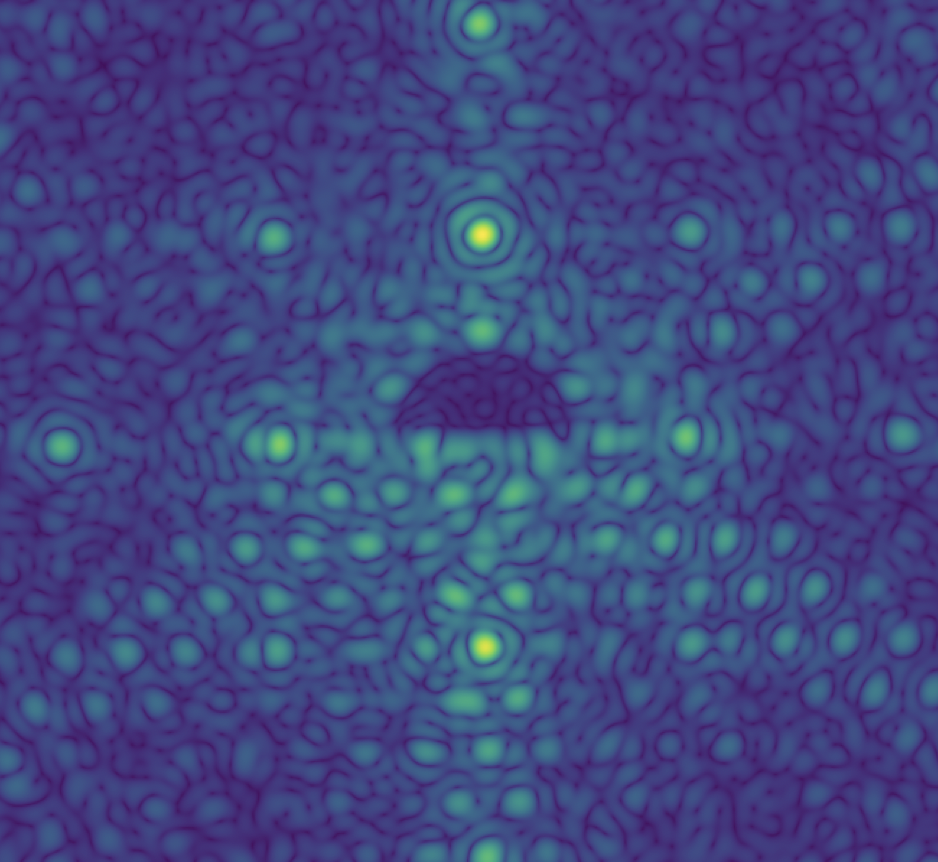}
   \end{tabular}
   \end{center}
   \caption[example] 
   { \label{fig:dark hole} 
Coronagraphic PSF before (left) and after (right) using dark hole digging techniques in simulation.}
   \end{figure}

Although we still have more to understand about our contrast limitations, SEAL now has coronagraphic abilities which allows us to use contrast as a metric for all wavefront sensing and control techniques developed on this testbed. Our future steps would be to direct our wavefront sensing towards the focal plane, through techniques such as dark hole digging~\cite{2007AAS...21113520G} (Figure~\ref{fig:dark hole}). The coronagraphic branch can also be further characterized with AO-corrected atmospheric turbulence, using the spatial light modulator (SLM) on SEAL. 

\acknowledgments 
 
Author Ashai Moreno would like to thank the Lloyd Robinson Research Fellowship, the UC LEADS program, and the STEM Diversity team at UCSC for their support.  

\bibliography{report} 
\bibliographystyle{spiebib} 

\end{document}